\documentclass[10pt,twocolumn]{article}

\usepackage{usenix-exp}
\usepackage{amsmath,amssymb}
\usepackage{float}
\usepackage{algorithm}
\usepackage{algpseudocode}
\usepackage{multirow}
\usepackage{url}

\newcommand{\sysname}{\textsc{Themis}}

\title{An Open-Source End-to-End FHE Implementation for Privacy-Preserving Llama 3 8B Inference\\[0.4em]
  }
\paperauthor{Yuhang Fan, Yusi Chen, Kanyu Ye, Zhuoran Ji}
\paperinstitute{School of Cyber Science and Technology, Shandong University;}

\begin{document}
\maketitle

\begin{abstract}

As large language model services increasingly move to the cloud, users typically send their prompts to model
providers, thereby raising the risk of privacy leakage.
Fully homomorphic encryption (FHE) enables a server to perform inference without decrypting the input, but
representing data as ciphertexts and performing homomorphic operations on them introduce additional storage and
computational overhead.
In CKKS-based LLM inference, the packing scheme determines how logical tensors are mapped to ciphertexts and
slots. It therefore determines the ciphertext count and the homomorphic-operation overhead of linear layers,
while constraining how data are represented and converted as they pass between linear layers, attention, and
nonlinear computation.
As model size and sequence length increase, inefficient layouts cause the overhead of encoding, computation, and
layout conversion to accumulate.

We present \sysname{}, an FHE inference system that co-designs ciphertext packing and model execution for Llama.
Starting from the weight-encoding bottleneck exposed by a THOR-style baseline, we design the Themis packing scheme.
Themis uses a feature-major cross-layer layout to unify residual connections and layer interfaces, and constructs
transient intra-operator layouts that match the computation patterns of linear projections and attention.
This design reduces redundant plaintext encoding of weights in wide projections. Within attention, $QK^\top$
produces scores that Softmax can consume directly, and $PV$ directly consumes the resulting probabilities,
avoiding intermediate repacking between Softmax and $PV$.
For nonlinear computation, we use minimax polynomial approximation over input intervals under a maximum
approximation-error constraint. We combine input-range control with joint error allocation guided by model quality
to reduce polynomial degree and multiplicative depth.
To the best of our knowledge, \sysname{} is the first open-source end-to-end implementation of Llama-3 using
GPU-accelerated CKKS.

With Llama-3-8B weights and a 128-token input, \sysname{} sequentially evaluates all 32 Transformer layers on a
single NVIDIA H100 80~GB GPU. Server-side end-to-end FHE evaluation takes 366.4~s and uses 58.9~GiB of peak
device memory.
With the same model, input, CKKS parameters, and hardware configuration, THOR's online FHE evaluation takes
1651.9~s, yielding a $4.51\times$ speedup.

\end{abstract}

\section{Introduction}
\label{sec:intro}

The Transformer architecture has advanced natural language processing and other sequence-modeling
tasks~\cite{vaswani2017attention}. As model sizes and context lengths grow, the computation and GPU memory
required for inference also increase. Large language models are therefore commonly deployed and served on cloud
servers. To access cloud computing resources, users must send their prompts to the service provider. When a prompt
contains personal, medical, commercial, or internal organizational information, the service provider has direct
access to the inference input. Protecting user privacy while retaining access to cloud-hosted models has thus
become a central problem in private LLM inference.

Existing private Transformer inference systems mainly use secure multi-party computation (MPC) or hybrid
cryptographic protocols. Systems such as Iron, PUMA, BOLT, and BumbleBee combine secret sharing, oblivious
transfer, and homomorphic encryption components to perform Transformer or LLM inference without revealing the
complete input to any single computing party~\cite{iron2022,puma2023,bolt2024,bumblebee2025}. These systems
support relatively large models, but they generally require a client or multiple servers to participate in
multi-round protocols and transmit large volumes of intermediate data during inference. Their security also relies
on protocol assumptions such as non-collusion among the participating parties. Communication volume and online
interaction therefore constitute the main deployment costs of MPC-based systems.

Fully homomorphic encryption (FHE) provides an alternative computation model. After a client encrypts its input,
the server can compute directly over the ciphertexts and return only a ciphertext result for the client to
decrypt~\cite{cheon2017ckks}. This model does not require the client to participate in every operator during
inference, nor does it require servers to maintain secret shares.
However, evaluating a Transformer under FHE requires expressing its computation using homomorphic additions,
multiplications, and rotations. The ciphertext layout also determines how data pass from one operator to the next.
When an operator's output layout differs from the input layout expected by its successor, additional homomorphic
operations are required for the conversion. Ciphertext packing is therefore a fundamental design choice: it
affects the numbers of ciphertexts and operations required for linear computation, determines how weights and
plaintexts are encoded, and determines the cost of layout conversions between operators. As the hidden dimension,
sequence length, and number of layers increase, the ciphertext count and the overhead of weight encoding and
layout conversion determined by the packing scheme increase accordingly. Packing design is therefore a key factor
in the efficiency of large-scale FHE LLM inference.

\subsection{Related Work}
Early private Transformer systems primarily use interactive MPC or hybrid HE--MPC protocols. Iron proposes
secure matrix multiplication based on compact HE packing and specialized protocols for Softmax, GELU, and
LayerNorm~\cite{iron2022}. PUMA extends two-party secure inference to LLaMA-7B through high-precision
approximations of GeLU and Softmax, along with secure Embedding and LayerNorm protocols~\cite{puma2023}.
BOLT further co-optimizes secure matrix multiplication and nonlinear computation~\cite{bolt2024}, while
BumbleBee designs low-communication matrix multiplication and activation-function protocols for large
Transformers~\cite{bumblebee2025}. These systems support relatively large models while protecting client inputs
and server models, but their online phases still require continuous communication between two parties.

FHE-only private inference has gradually expanded from individual operators to complete Transformers. NEXUS
proposes several homomorphic matrix operations and nonlinear approximations, but the composition of its different
packing formats and the costs of converting among them are not incorporated into an end-to-end execution
path~\cite{nexus2025}. THOR uses diagonal-major encoding and compact ciphertext packing to perform end-to-end
FHE inference for BERT-base on a single GPU~\cite{thor}. MOAI further reduces layout conversions and rotations by
coordinating packing formats across modules~\cite{moai2025}. Recent work has extended CKKS inference to
Llama-2-7B and supports longer inputs by distinguishing public tokens from encrypted
tokens~\cite{park2026llama}. These advances show that the performance of an FHE Transformer depends not only on
individual homomorphic operators, but also on whether the packing scheme can accommodate the model scale and
remain compatible throughout the computation pipeline. After adapting a THOR-style layout to Llama-3-8B, we
further observe that the online encoding of a large number of plaintext weights becomes the main bottleneck in
linear computation. Section~\ref{sec:baseline-bottleneck} presents the corresponding timing breakdown.

Nonlinear computation constitutes another major source of cost. Because CKKS natively supports approximate
arithmetic, RMSNorm, Softmax, and SiLU generally must be converted into polynomial or iterative circuits. Starting
from a doubling-iteration formula for Softmax, Cho et al. propose a fast method for homomorphic
Softmax~\cite{hosomax}. Park et al. further apply a prepended attention-sink
token~\cite{xiao2024streamingllm} and the orthogonal rotations of QuaRot~\cite{ashkboos2024quarot} to FHE
Llama to mitigate the expansion of approximation intervals caused by attention sinks and dimensional
outliers~\cite{park2026llama}. These methods provide executable components for different nonlinear functions, but
a complete model still requires the input ranges, polynomial errors, and CKKS precision to be determined jointly
so that local approximations preserve model quality after propagation through multiple layers.

\subsection{Our Approach and Contributions}

We present \sysname{}, an FHE inference system that co-designs ciphertext packing and model execution for Llama.
We first construct a THOR-style Llama-3 baseline and identify the weight-encoding bottleneck through a stage-level
timing breakdown of this baseline. Based on this observation,
Themis fixes a feature-major layout as the hidden-state interface for residual connections and inter-layer transfer,
while using short-lived intra-operator layouts for linear projections and attention. A structured weight
representation allows encoded weight diagonals to be reused across multiple input ciphertexts, reducing redundant
weight encoding.

Within attention, $QK^\top$ produces scores in a representation that Softmax can consume directly. Softmax
preserves this representation for its output probabilities, which are then consumed directly by $PV$. This
dataflow avoids intermediate repacking between Softmax and $PV$. The system also applies range control and
polynomial approximation to RMSNorm, Softmax, and SiLU, and jointly searches for candidate approximation-error
budgets through model-level error injection.
Together, these components form a server-side FHE execution path through all Transformer layers of Llama-3.

Our main contributions are as follows:
\begin{itemize}[leftmargin=1.1em,itemsep=1pt,topsep=2pt]
  \item \textbf{The Themis packing scheme.} We propose an FHE Llama packing scheme for both linear layers and
        attention. Themis organizes a complete block with a stable cross-layer layout and short-lived intra-operator
        layouts. It reuses encoded weight diagonals in the wide projections of Llama-3. Within attention,
        $QK^\top$ produces scores that Softmax can consume directly, and $PV$ directly consumes the resulting
        probabilities, avoiding intermediate repacking between Softmax and $PV$.
  \item \textbf{Nonlinear computation guided by full-model quality.}
        Starting from a uniform 12-bit precision, we use model-level error injection to compare the sensitivity of
        RMSNorm, Softmax, and SiLU to different error magnitudes and jointly search for candidate error
        configurations for these operators. The candidate budgets and input ranges jointly determine the degrees
        of the minimax polynomials, whose effects on model quality are then evaluated separately.
  \item \textbf{An open-source end-to-end FHE implementation of Llama-3.} We implement and open-source the
        complete model-conversion pipeline, GPU ciphertext execution, and experimental configurations.
        To the best of our knowledge, this is the first open-source end-to-end FHE implementation of Llama-3. Using the actual
        Llama-3-8B weights and a 128-token input, the system sequentially evaluates all 32 Transformer layers on a
        single NVIDIA H100 80~GB GPU, taking 366.4~s for server-side end-to-end FHE evaluation and using
        58.9~GiB of peak device memory.
\end{itemize}

\section{Background}
\label{sec:background}
\subsection{The CKKS Computation Model}
\label{sec:ckks-background}
CKKS~\cite{cheon2017ckks} is an RLWE-based fully homomorphic encryption scheme for approximate numerical
computation. CKKS encodes a vector of real or complex values into a plaintext polynomial, which can then be encrypted into a
ciphertext. Decryption followed by decoding recovers an approximation to the original vector. A ciphertext can
pack multiple values into slots, and homomorphic operations act on these slots in SIMD fashion. Ciphertext
multiplication followed by rescaling progressively lowers the ciphertext level. Bootstrapping refreshes a
ciphertext and restores its level budget, allowing evaluation to continue.

We use the following primitive CKKS operations. $\mathsf{Encode}(\mathbf{m})$ returns a plaintext polynomial
whose slots represent the numerical vector $\mathbf{m}$. $\mathsf{Add}$, $\mathsf{PCMult}$, and $\mathsf{Mult}$
compute ciphertext addition,
slot-wise plaintext--ciphertext multiplication, and slot-wise ciphertext--ciphertext multiplication, respectively.
$\mathsf{Rot}_k$ cyclically shifts the slot vector by $k$ positions. $\mathsf{KeySwitch}$ transforms a ciphertext
representation using an evaluation key and underlies rotations and relinearization after ciphertext
multiplication. $\mathsf{Rescale}$ adjusts the scaling factor and ciphertext modulus after
multiplication while reducing the level. $\mathsf{Bootstrap}$ refreshes a ciphertext and restores its level budget,
but incurs high latency and additional approximation error.

\subsection{The Llama-3-8B Architecture}
\label{sec:llama-background}
Llama-3 uses a decoder-only Transformer architecture~\cite{vaswani2017attention,dubey2024llama3}. Compared
with a standard Transformer decoder, each Llama-3 block uses RMSNorm-based pre-normalization, encodes the
positional information of $Q$ and $K$ with RoPE, and employs grouped-query attention (GQA) and a SwiGLU FFN.
GQA allows multiple query heads to share key/value heads, thereby reducing the output dimensions of the K/V
projections and the size of the KV cache. SwiGLU replaces the single activation path in a standard FFN with a
SiLU-gated branch.
%


In our FHE execution model, the server holds the model weights in plaintext, while model activations remain
encrypted. Consequently, the Q/K/V/O and gate/up/down projections become plaintext--ciphertext matrix
multiplications, whereas $QK^\top$ and $PV$ multiply two encrypted operands. RMSNorm, Softmax, and SiLU must be
converted into polynomial or iterative arithmetic circuits supported by CKKS. This division gives rise to the
linear-packing and nonlinear-approximation problems discussed in Section~\ref{sec:baseline}.


\section{FHE Llama-3: Problem Definition and Baselines}
\label{sec:baseline}

\subsection{Private Inference and Threat Model}
\label{sec:baseline-pipeline}
We consider a private inference setting consisting of a client and a cloud server. The client holds the prompt and
the CKKS secret key, prepares the input locally, and sends the encrypted input and the evaluation keys required by
the protocol to the server.
The server holds the model weights in plaintext, performs Llama-3 inference over
ciphertexts, and returns the result ciphertext to the client for decryption. Apart from submitting the input and
receiving the result, the client does not participate in server-side homomorphic evaluation. The input and all
prompt-dependent intermediate states remain encrypted on the server. The client tokenizes the prompt, performs the
embedding lookup, and encrypts the resulting input before transmission. These client-side preprocessing and
encryption steps are excluded from FHE evaluation time.

We assume an honest-but-curious server: it follows the prescribed inference protocol but may use the input
ciphertexts, evaluation keys, intermediate ciphertexts, and observable public information to infer the client
input. Under the RLWE security assumption underlying CKKS, our privacy objective is to protect the prompt and its
derived intermediate states. The model architecture and weights, sequence length, cryptographic parameters, and
runtime are considered public information. We do not consider a malicious server that deviates from the protocol,
side-channel attacks, denial of service, or access-pattern leakage, nor do we protect the server's model weights.
Under this system and security boundary, the following subsections describe server-side nonlinear computation,
linear-packing baselines, and their performance bottlenecks.

\subsection{Nonlinear Computation}
\label{sec:baseline-transform}

\subsubsection{Polynomial Approximation}
\label{sec:baseline-approximation}

CKKS can directly evaluate additions and multiplications. RMSNorm, SiLU, and Softmax contain functions such as
inverse square root, exponentiation, and division that CKKS cannot evaluate directly. These functions must be
replaced by polynomial or iterative arithmetic circuits. For a function $f$ defined over an interval
$I=[a,b]$, let $p_d$ denote a polynomial of degree at most $d$. The uniform absolute error is defined as
\begin{equation}
  E_d(f;I)=\min_{\deg(p)\leq d}\max_{x\in I}|f(x)-p(x)|.
  \label{eq:uniform-approximation}
\end{equation}
Given a maximum allowable error $\tau$, the minimum required degree is
\begin{equation}
  d_\tau(f;I)=\min\{d:E_d(f;I)\leq\tau\}.
  \label{eq:required-degree}
\end{equation}
Given an approximation interval $I$ and a candidate degree, the Remez algorithm generates an approximate minimax
polynomial.
We then search over candidate polynomial degrees and use the error threshold $\tau$ to determine the smallest degree
that satisfies the requirement. The approximation interval and error threshold are therefore the two central factors
that determine the polynomial degree.

This dependence can be described by a standard approximation bound. Let $h=(b-a)/2$, and assume that $f$ has an
$r$th derivative over $I$ and that $f^{(r)}$ satisfies an $\alpha$-H\"older condition. Define $s=r+\alpha$.
After scaling $I$ to $[-1,1]$, a Jackson-type bound gives~\cite{devore1993constructive}
\begin{equation}
  E_d(f;I) \leq C_f h^s d^{-s}.
  \label{eq:degree-power-law}
\end{equation}
Here, $C_f$ depends on the smoothness of the function over the interval. A sufficient degree for achieving error
$\tau$ therefore satisfies
\begin{equation}
  d \geq h\left(\frac{C_f}{\tau}\right)^{1/s}.
  \label{eq:degree-range-precision}
\end{equation}
Under the same assumptions on function regularity, this bound shows that the degree grows linearly with the
interval length and grows as $\tau^{-1/s}$ as the allowable error becomes tighter. It is a sufficient bound rather
than a strict equality for all functions: the actual minimum degree also depends on the function shape and the
coefficient-generation method. Analytic functions admit faster exponential convergence, but interval scaling
still changes the corresponding constants and convergence parameters. We therefore treat the approximation
interval and maximum error jointly as configuration parameters for each nonlinear operator.

FHE nonlinear circuits require two distinct design choices: how to cover the input domain and how to compose
arithmetic subroutines. For the former, piecewise methods partition a wide domain using comparisons and masks and
approximate each segment separately. When calibration can determine a finite interval, each scalar function can
instead be approximated directly by a minimax polynomial over that interval. Iterative identities such as
Newton--Raphson, Goldschmidt, and Softmax doubling may then compose these scalar approximations; they are not
alternatives to interval selection. We avoid comparison-based domain partitioning and directly approximate each
scalar subfunction over a calibrated interval. Softmax composes these polynomials within a doubling iteration.

We also compare the input distributions of different nonlinear functions during inference on WikiText-2 and C4. Across corresponding
operator positions in all 32 layers, the median C4-to-WikiText-2 ratio of p99 magnitudes is 1.0, and we observe no
overall scale shift between the two general-language corpora. Some layer--operator positions nevertheless deviate.
This result supports using a calibration set to initialize approximation intervals, but does not guarantee that
all inference inputs remain within them. Section~\ref{sec:model-preprocess} therefore adds safety margins and
out-of-range checks.

\subsubsection{Input-Range and Outlier Mitigation}
\label{sec:baseline-range}
Intermediate activations in LLMs generally contain a small number of outliers whose magnitudes are much larger
than those of the remaining elements. These outlying coordinates expand the common approximation interval used
for all inputs, causing every slot to incur the evaluation of a higher-degree polynomial. A larger dynamic range
also increases the precision requirement of CKKS bootstrapping. Reducing the maximum magnitude can therefore
reduce the polynomial degree required to meet the same error target.

To the best of our knowledge, Scaling up Privacy-Preserving ML~\cite{park2026llama} is the first work to combine
two types of outlier-mitigation methods originally developed for ordinary LLM inference in an FHE Llama setting:
attention sinks are handled along the token dimension, and dimensional outliers are handled along the feature
dimension. The first type of outlier is caused by attention sinks formed by special tokens during the initial stage
of inference, whose magnitudes can be much larger than those of subsequent content tokens. That work prepends a
fixed sink prefix to the input and precomputes its key--value states. Because the prefix is independent of the user
query, it can be deployed as a public static cache, avoiding the encryption and online-evaluation costs for these
tokens. The purpose of this preprocessing is to stabilize the activation range of subsequent tokens rather than
to change the definition of Softmax~\cite{xiao2024streamingllm,park2026llama}.

The second type of outlier appears when a small number of feature coordinates consistently exhibit large
magnitudes across multiple layers. QuaRot~\cite{ashkboos2024quarot} uses computational invariance to insert
random orthogonal Hadamard rotations into hidden states and adjacent weights, distributing magnitudes concentrated
in a small number of coordinates across multiple coordinates. Where they can be absorbed algebraically, the
rotations can be fused offline into public weights and thus add no homomorphic operations to the online FHE
circuit. Scaling adapts this rotation-based processing to FHE Llama and combines it with the prefix method to
control nonlinear input ranges~\cite{park2026llama}. The two methods act on the token and feature axes,
respectively, and cannot replace each other. Their common goal is to reduce the dynamic range required by
polynomial approximation and bootstrapping.

We adopt this FHE adaptation as the model-preprocessing baseline and use public model weights with the QuaRot
transformations fused offline.

\subsubsection{Approximation-Error Budget}
\label{sec:baseline-error-budget}
Existing FHE LLM implementations generally begin by selecting a uniform conservative precision for approximation
errors. Through plaintext error simulation, Park et al. observe that a 12-bit error budget does not degrade PPL
relative to their FP16 baseline, and use 12 bits as the target precision for subsequent
configurations~\cite{park2026llama}. We adopt this setting and set the maximum approximation errors of the
RMSNorm inverse square root, SiLU, and the initial exponential and normalization inverse-square-root stages of
Softmax to $\tau_0=2^{-12}$. This uniform setting provides a reproducible starting point, but also leaves different
amounts of precision margin for different nonlinear operators.

If a separate tolerance $\tau_i^{(\ell)}$ is assigned to every nonlinear function in every layer, error allocation
requires a search over a continuous high-dimensional parameter space. Different operators and layers also have
different sensitivities to model quality, so an appropriate configuration cannot be derived directly from the
uniform 12-bit starting point. To reduce the search complexity, we initially ignore differences among layers for
operators of the same type and assign shared operator-level tolerances separately to the two RMSNorms, Softmax,
and SiLU. This simplification transforms layer-wise error allocation into allocation among four classes of
nonlinear operators.


Based on this simplification, we divide calibration into a single-operator search and a joint search. The
single-operator search keeps all other operators exact and compares the sensitivity of different operators and
error magnitudes through bounded error injection. A uniform-tolerance search first identifies a feasible common
candidate under the injection criterion; the 12-bit configuration is used as a conservative reference rather than
as a mandatory starting point. The joint search then adjusts operator-level tolerances one at a time, according to a
predetermined priority, while perturbations in all four nonlinear classes are present. This procedure screens
candidate approximation budgets. Because injected perturbations are only a proxy for sensitivity, they do not
guarantee that the corresponding Remez polynomials produce the same PPL; model quality must be evaluated
separately after the final polynomials are generated. Section~\ref{sec:nonlinear-calibration} describes the search
method and error-interaction analysis.

Relaxing the error budget can reduce not only the corresponding homomorphic multiplications but also affect the
bootstrapping schedule. Bootstrapping positions are jointly determined by the level requirements of the complete
execution chain and the depths of the operators. Reducing the degree of RMSNorm or Softmax therefore does not
necessarily reduce the number of bootstrapping operations. In contrast, SiLU is located on the FFN gating branch,
which has a width of $14336$. In a single-operator pilot test, we keep the approximation interval unchanged and
relax the SiLU tolerance from $2^{-12}$ to $3.82\times10^{-3}$. The sum of the SiLU polynomial degrees across
32 layers decreases from $2048$ to $998$, allowing the subsequent FFN linear layer to use fewer limbs under the
current schedule. This makes SiLU the highest-priority target in the current error budget.

\subsection{Linear Computation and Packing Baselines}
\label{sec:baseline-linear}

\subsubsection{Matrix Representations and Linear Computation in CKKS}
CKKS slot encoding maps a vector to the SIMD slots of a plaintext polynomial, such that homomorphic addition and
multiplication correspond to slot-wise addition and multiplication, respectively~\cite{cheon2017ckks}. For a slot
vector $x\in\mathbb{R}^{D}$ and a matrix $A\in\mathbb{R}^{D\times D}$ (zero-padded when necessary), define
$D_k(A)[i]=A[i,(i+k)\bmod D]$ as the $k$th cyclic diagonal. Let $\rho^k$ be the left rotation satisfying
$(\rho^k(x))_i=x_{(i+k)\bmod D}$. Matrix--vector multiplication can then be written as
\begin{equation}
  Ax=\sum_{k=0}^{D-1}D_k(A)\odot\rho^k(x),
  \label{eq:slot-diagonal-matvec}
\end{equation}
where $\rho^k$ denotes a slot rotation. Row, column, and diagonal packing specify how a logical matrix is laid out
in slots. BSGS is an evaluation strategy that reuses some rotations over a given layout rather than an independent
packing scheme. A slot representation can directly support element-wise polynomials, RoPE, and masks. The
cross-slot reductions required by Softmax and RMSNorm can also be implemented using rotations and accumulation,
making it convenient for passing ciphertexts between linear and nonlinear layers.

However, slot encoding alone does not define a complete linear algorithm. An executable scheme must also specify
the token--feature mapping, organization of plaintext weights, rotation set, cross-ciphertext reductions, and
output layout. MOAI~\cite{moai2025} demonstrates the importance of layout compatibility: its projection results
retain column packing, $QK^\top$ produces diagonal packing, Softmax preserves this representation, and the result
is then multiplied by column-packed $V$ before column packing is restored. If each operator selects its layout
independently, the same conversions require additional rotations, masks, or repacking. The cost of slot packing
therefore depends on the lifetime of a layout throughout the operator chain and cannot be assessed solely from an
individual matrix-multiplication kernel.

Another class of methods uses coefficient encoding, which writes real values directly into the coefficients of a
plaintext polynomial. Bae et al. organize coefficient-encoded RLWE/MLWE ciphertexts as matrix equations, reduce
wide PCMM to a small number of plaintext matrix multiplications, and use small-ring representations and BLAS
implementations~\cite{bae2024pcmm}. Scaling up Privacy-Preserving ML uses this method for the QKV, output, and
FFN projections in Llama while retaining slot encoding for other operations~\cite{park2026llama}. SlotToCoeffs
and CoeffsToSlots between the two representations are homomorphic DFTs and involve level consumption, bit
reversal, and RLWE/MLWE layout adjustments. These conversions incur corresponding computation and limb costs.
EasyFHE does not yet provide a complete execution path for these representation conversions. We therefore discuss
coefficient encoding as an alternative design point, but restrict the subsequent evaluation to THOR and Themis,
both of which operate in the slot domain.

\subsubsection{THOR Packing}

THOR~\cite{thor} targets non-interactive Transformer inference for BERT-base and organizes
plaintext--ciphertext matrix multiplication (PC-MM) and ciphertext--ciphertext matrix multiplication (CC-MM)
around cyclic-diagonal representations. For $A\in\mathbb{R}^{n\times m}$ and
$B\in\mathbb{R}^{m\times n}$, where $n\geq m$ and $m\mid n$, let $U_k(A)$ and $L_k(B)$ denote the $k$th upper
and lower cyclic diagonals of the matrices, respectively. A lower diagonal of the product satisfies
\begin{equation}
  L_r(AB)=\sum_{\ell=0}^{m-1}\rho^r\!\left(U_{\ell-r}(A)\right)\odot L_\ell(B),
  \label{eq:thor-diagonal-product}
\end{equation}
where the diagonal indices wrap around according to the corresponding dimension. This identity transforms matrix
multiplication into rotations, slot-wise multiplications, and accumulations of diagonal vectors, while preserving
the diagonal semantics of the output.

When a single matrix diagonal is shorter than the CKKS slot capacity, encrypting each diagonal separately leaves
many slots unused. THOR's multi-diagonal batched encoding first interleaves corresponding diagonals from multiple
attention heads or parallel submatrices and then concatenates $c$ adjacent batched diagonals in one ciphertext.
Here, $c$ is called the diagonal packing capacity. This layout exploits both inter-matrix parallelism and the
remaining slot capacity, providing a unified diagonal input for subsequent PC-MM and CC-MM.

For linear projections, THOR divides public weights and encrypted activations into smaller square blocks and
computes in parallel the submatrix products located on the same block diagonal. It further uses a transposed data
flow, for example by computing $Q^\top=W_Q^\top X^\top$, to absorb some rotations that would otherwise act on
input ciphertexts into the preprocessing of public plaintext weight diagonals. For $QK^\top$ and $PV$ in
attention, THOR first organizes the inputs into replicated diagonals required by lower--lower multiplication. A
BSGS reformulation then defers some rotations until after ciphertext multiplication and local accumulation, which
allows these rotations to be merged. The diagonal representation allows row-wise Softmax and LayerNorm to be
evaluated using the corresponding upper-diagonal semantics. However, the complete workflow still includes
transposition, replication, masking, and format adjustments between upper and lower diagonals, and not all
operators can be connected without conversion~\cite{moai2025}.

THOR's original analysis primarily optimizes slot utilization and the numbers of rotations, multiplications, and
key-switching operations. Its PC-MM interface takes encoded and internally rotated plaintext weight diagonals as
input, but does not discuss how these plaintexts are generated, cached, and reused across projection blocks and
computation batches. For the $4096$-dimensional hidden state and $14336$-dimensional FFN of Llama-3, the lifetime
of these plaintexts becomes a system issue that must be analyzed separately. The next section therefore separates
the costs of weight preparation, rotations, and multiplications in the linear path and identifies the primary
bottleneck when scaling the THOR baseline to Llama-3.

\subsection{The Encoding Bottleneck at Llama-3 Scale}
\label{sec:baseline-bottleneck}
To identify the primary performance bottleneck after scaling THOR packing to Llama-3, we profile Layer~16 of
Llama-3-8B with a 128-token input and break down the runtime of each component.
Table~\ref{tab:thor-layer-breakdown} lists operator-level runtimes in execution order.

\begin{table}[H]
  \centering
  \caption{Median runtime breakdown over three runs for one Llama-3-8B layer using THOR packing}
  \label{tab:thor-layer-breakdown}
  \scriptsize
  \setlength{\tabcolsep}{3pt}
  \begin{tabular}{c p{0.55\columnwidth} r}
    \toprule
    Module & Operation & Time (s) \\
    \midrule
    RMSNorm$_1$ & Normalization & 0.05 \\
    \midrule
    \multirow{5}{*}{Attention} & QKV projection & 5.44 \\
     & $QK^\top$ & 4.83 \\
     & Softmax & 4.82 \\
     & $PV$ & 1.95 \\
     & $O$ projection & 5.39 \\
    \midrule
    RMSNorm$_2$ & Normalization & 0.05 \\
    \midrule
    \multirow{4}{*}{FFN} & gate projection & 11.69 \\
     & up projection & 8.00 \\
     & SiLU & 0.40 \\
     & down projection & 7.50 \\
    \midrule
    \textbf{Linear computation} & QKV, $O$, FFN projections, $QK^\top$, and $PV$ & \textbf{44.80} \\
    \textbf{Nonlinear operators} & RMSNorm, Softmax, and SiLU & \textbf{5.32} \\
    Other FHE operations & Bootstrapping, RoPE, layout conversion, and unassigned runtime & 9.12 \\
    \textbf{Total} & Complete single-layer FHE evaluation & \textbf{59.24} \\
    \bottomrule
  \end{tabular}
\end{table}
A Transformer layer in Llama-3 contains both plaintext--ciphertext and ciphertext--ciphertext linear computation.
The QKV, $O$, and FFN projections perform plaintext--ciphertext multiplication between model weights and
ciphertexts, while $QK^\top$ and $PV$ in attention perform ciphertext--ciphertext multiplication. In
Table~\ref{tab:thor-layer-breakdown}, the former takes $38.02$~s and the latter takes $6.78$~s. Together they take
$44.80$~s, accounting for $75.6\%$ of the single-layer runtime. Linear computation is therefore the primary cost of this baseline.

\begin{table}[H]
  \centering
  \caption{Median runtime breakdown over three runs for Llama-3-8B linear computation using THOR packing}
  \label{tab:thor-linear-ops}
  \scriptsize
  \setlength{\tabcolsep}{2pt}
  \begin{tabular}{lrrrrrrr}
    \toprule
    Scope & \shortstack{Weight\\pack/enc.} & \shortstack{Aux.\\enc.} & Rot. & Mult. & Merge & Rescale & Total \\
    \midrule
    \multicolumn{8}{l}{\emph{Plaintext--ciphertext multiplication}} \\
    $Q$ & 2.91 & -- & 0.04 & 0.28 & 0.42 & 0.06 & 3.70 \\
    $K+V$ & 1.23 & -- & 0.04 & 0.14 & 0.10 & 0.03 & 1.53 \\
    $O$ & 2.47 & -- & 0.46 & 0.27 & 1.50 & 0.19 & 4.88 \\
    FFN & 21.34 & -- & 0.17 & 2.70 & 1.68 & 0.66 & 26.55 \\
    \midrule
    \multicolumn{8}{l}{\emph{Ciphertext--ciphertext matrix multiplication}} \\
    $QK^\top$ & -- & 0.48 & 3.88 & 0.22 & 0.26 & 0.08 & 4.94 \\
    $PV$ & -- & 0.37 & 1.64 & 0.18 & 0.22 & 0.00 & 2.40 \\
    \midrule
    PC subtotal & 28.02 & -- & 0.70 & 3.40 & 3.71 & 0.94 & 36.76 \\
    CC subtotal & -- & 0.85 & 5.52 & 0.41 & 0.48 & 0.08 & 7.34 \\
    All linear comp. & \textbf{28.02} & 0.85 & 6.22 & 3.80 & 4.19 & 1.02 & 44.10 \\
    \bottomrule
  \end{tabular}
\end{table}

Table~\ref{tab:thor-linear-ops}  breaks down the two types of linear computation through an independent
operator-level profile. The two types of multiplication take $44.10$~s in total. Weight packing and encoding for
plaintext--ciphertext multiplication take $28.02$~s, or $63.5\%$ of this time. Lower-level timers further divide
this cost into $10.99$~s for packing and $17.03$~s for encoding. Encoding is the largest individual operation,
accounting for $38.6\%$ of linear-computation time. Organizing and encoding weight plaintexts is therefore the
primary bottleneck in the current baseline.

This bottleneck arises from a mismatch between the reuse granularity of the weight representation in our
THOR-style Llama-3 adaptation and the model's wide projections. Designed for BERT, THOR treats encoded plaintext
weight diagonals as preprocessed inputs to PC-MM and focuses primarily on slot utilization and the costs of online
rotations and multiplications~\cite{thor}. In the page-diagonal representation used by our adaptation, a plaintext
weight represents not only a logical diagonal but is also bound to a particular input page, output group, real or
imaginary component, and rotation copy. An encoding result can be reused directly only when all of these
physical-layout attributes are identical. After this representation is scaled to a $4096$-dimensional hidden state
and a $14336$-dimensional FFN, the combinations of layout attributes produce $425{,}984$ plaintext weight objects
per layer. Pre-encoding and caching these objects incur corresponding storage and residency costs, whereas
generating them on demand repeatedly performs packing and encoding at runtime. The root cause is therefore that
this THOR-style layout cannot readily amortize the encoding cost of the same logical weight across multiple
compatible input ciphertexts. Themis instead organizes plaintexts by physical weight page and logical diagonal so
that encoding results can be reused across token-shard inputs. Section~\ref{sec:themis} describes its layout and
matrix multiplication in detail.

\section{Themis: A New Packing Scheme for FHE LLMs}
\label{sec:themis}


\subsection{Design Goals and Overview of Themis}
\subsection{Cross-Layer Feature-Major State and Intra-Operator Layouts}
\label{sec:themis-layouts}
\subsection{Packing-Aware Homomorphic Linear Computation}
\label{sec:themis-matmul}
\subsection{\texorpdfstring{C--$\Delta$}{C--Delta} Attention Layout and Computation}
\label{sec:themis-attention}
\subsection{Layout Transitions in a Llama-3 Block}
\label{sec:themis-llama-linear}
\subsection{Complexity and Overhead Analysis}
\label{sec:themis-complexity}
\section{System Implementation of FHE Llama}
\label{sec:themis-system}

Section~\ref{sec:themis} defines Themis's layouts and the principles of its linear computation. This section describes
the process of transforming a public Llama-3 model into an FHE execution graph. We first transform the weights
and calibrate the input ranges of nonlinear functions, generate the nonlinear approximations and numerical
configurations, connect each projection layer to the homomorphic linear kernels, and finally assemble a layer-wise
execution graph. The resulting artifacts include transformed weights, measured input intervals, polynomial
coefficients, and bootstrapping positions. Section~\ref{sec:evaluation} evaluates correctness, model quality, and
performance after these artifacts have been fixed.

\subsection{Model Preprocessing and Input-Range Calibration}
\label{sec:model-preprocess}

Section~\ref{sec:baseline-range} describes how QuaRot mitigates feature outliers. We apply the QuaRot
transformation to the public Llama-3 weights and fuse the Hadamard transformations preceding the $O$ and down
projections into adjacent weights, so they add no homomorphic computation to FHE execution. The transformed
model is used consistently for subsequent interval measurement, polynomial generation, and FHE linear
computation, ensuring that the model used during calibration is identical to the model used during actual
execution.

Section~\ref{sec:baseline-approximation} compares the distributions of intermediate values at the same layers and
nonlinear positions on WikiText-2 and C4. Their p99-magnitude ratios are close to $1$, and no clear scale shift is
observed between the two corpora. To avoid potential out-of-bound inputs, however, we define the approximation
interval for each nonlinear input in each layer by multiplying the observed extrema by a predetermined expansion
factor. SiLU uses a symmetric interval based on the maximum absolute value. The input to the inverse square root
approximated in RMSNorm is positive, so a positive interval is used. The approximation intervals for Softmax are
described separately in its implementation. The safety margin reduces the risk of out-of-bound inputs, but also
increases the polynomial degree for a given error target. The current configuration uses an expansion factor of
$1.2$.

\subsection{Nonlinear-Function Approximation and Error Calibration}
\label{sec:nonlinear-calibration}
Section~\ref{sec:model-preprocess} determines nonlinear input intervals for each layer. This section further
determines the error budgets and converts them into executable polynomials. A theoretically optimal solution to
the original problem would require selecting a continuous error-budget value separately for RMSNorm$_1$,
Softmax, RMSNorm$_2$, and SiLU in each of the 32 layers. These budgets interact as errors propagate across
layers, and there is no directly solvable analytical relationship between model quality and the error budgets.
Obtaining a theoretically optimal configuration is therefore difficult. We retain layer-wise approximation
intervals but let operators of the same type share one output-perturbation budget across layers, reducing the search
to four coupled variables. RMSNorm and SiLU can each be implemented by a polynomial for a single target
function. Softmax must be decomposed into exponential, reduction, and inverse-square-root circuits, so its
error-injection positions must be defined separately.

\subsubsection{Decomposed Approximation of Softmax}

Unlike RMSNorm and SiLU, Softmax includes exponentiation, sum reduction, and normalization and cannot be
directly replaced by a single element-wise polynomial. Cho et al. propose a fast homomorphic Softmax algorithm
based on the Softmax doubling identity~\cite{hosomax}. We use this algorithm as the basic computational framework
for homomorphic Softmax.

The input to Softmax is an attention-score vector $x=(x_1,\ldots,x_n)\in\mathbb{R}^{n}$ of length $n$. Let
$m=\max_{1\leq i\leq n}x_i$, let $z=x-m\in[-M,0]$, and set
\begin{equation}
  k=\left\lceil \log_2 M-\log_2\ln n \right\rceil ,
  \label{eq:softmax-k}
\end{equation}
We first apply an exponential polynomial to $z/2^k$ to obtain $u^{(0)}=p_{\exp}(z/2^k)$. We then increase the
input scale in each round according to the Softmax doubling identity. After $k$ rounds, this yields
$\operatorname{Softmax}(z)=\operatorname{Softmax}(x)$:
\begin{equation}
  \operatorname{Softmax}(2x)_i
  =\frac{\operatorname{Softmax}(x)_i^2}
  {\sum_{j=1}^{n}\operatorname{Softmax}(x)_j^2}.
  \label{eq:softmax-doubling}
\end{equation}

The FHE implementation uses the statistical maximum $\widehat U$ and interval length $\widehat M$ obtained
during calibration in place of $M$ in Equation~\eqref{eq:softmax-k}. As long as $\widehat U$ covers the maximum
score in the row, the translation invariance of Softmax guarantees
$\operatorname{Softmax}(\widehat z)=\operatorname{Softmax}(x)$. An insufficient $\widehat U$ causes the input to
leave the approximation interval, whereas an excessively large upper bound expands the exponential interval and
increases $k$. Most layers share $\widehat U$ and $k$ within the layer. For positions with wider distributions,
such as Layer~31, we use a head-wise $\widehat U$ and set $k=\max_h k_h$ to maintain the same number of iterations
within a SIMD batch.

The implementation follows the wide- and narrow-track structure of the algorithm of Cho et al. The main track
holds all score coordinates and performs exponentiation, slot-wise multiplication, and squaring. The auxiliary
track holds only the reduction value $s^{(j)}$ for each score row and evaluates the inverse square root over a small
number of valid slots. This structure avoids repeatedly evaluating a high-degree inverse square root over all
score slots and confines bootstrapping (BS) to compact auxiliary-track ciphertexts. In each round, the
implementation first accumulates valid rows distributed across multiple ciphertexts into one ciphertext, then
computes $\lambda^{(j)}$ and performs BS on that ciphertext. It subsequently extracts the normalization factor for
each row and broadcasts it back to the main track. Algorithm~\ref{alg:fhe-softmax} gives the complete procedure.

\begin{algorithm}[t]
  \caption{Homomorphic Softmax using a public statistical shift}
  \label{alg:fhe-softmax}
  \begin{algorithmic}[1]
    \Require Score vector $x$, public parameters $\widehat U$ and $k$, exponential polynomial
    $p_{\exp}$, and inverse-square-root polynomial $p_{\mathrm{rsqrt}}^{(j)}$ for round $j$
    \Ensure Softmax approximation $u^{(k)}$
    \State $z\gets x-\widehat U$
    \State $u^{(0)}\gets p_{\exp}(z/2^k)$ \Comment{Slot-wise computation on the main track}
    \For{$j=1,\ldots,k$}
      \State $s^{(j)}\gets\sum_{i=1}^{n}(u_i^{(j-1)})^2$ \Comment{Reduce to the auxiliary track}
      \State $\lambda^{(j)}\gets p_{\mathrm{rsqrt}}^{(j)}(s^{(j)})$ \Comment{Auxiliary track}
      \State $u^{(j)}\gets(\lambda^{(j)}u^{(j-1)})^2$ \Comment{Slot-wise squaring on the main track}
    \EndFor
    \State \Return $u^{(k)}$
  \end{algorithmic}
\end{algorithm}

For the exponential approximation in Softmax, the above scaling restricts the input to $[-\ln n,0]$. The
inverse-square-root polynomial required for normalization in each round takes the reduction value $s^{(j)}$ as
input. The first-round reduction value $s^{(1)}=\sum_i (u_i^{(0)})^2$ satisfies
$s^{(1)}\in[1/n,n]$, while subsequent normalized rounds satisfy $s^{(j)}\in[1/n,1]$. The theoretical upper
bound grows with $n$, but actual Llama inputs generally cover only a narrower range. Direct approximation over
the theoretical interval increases the degree of the inverse-square-root polynomial. We therefore measure the
actual range of $s^{(j)}$ for each layer and iteration and generate the corresponding polynomial after
multiplying by a safety factor.

\subsubsection{Approximation-Error Proxy and Uniform Tolerance}
Let $\tau$ denote a candidate upper bound on the local approximation error, let
$\widetilde{\mathcal{M}}_{\tau}$ denote the plaintext model that independently injects
$e\sim\mathcal{U}[-\tau,\tau]$ at each approximation position, and let $\mathcal{M}_{0}$ denote the reference
model using exact nonlinear functions. We use
\begin{equation}
  \Delta\mathrm{PPL}_{\mathrm{inj}}(\tau)
  =\mathrm{PPL}(\widetilde{\mathcal{M}}_{\tau})-
   \mathrm{PPL}(\mathcal{M}_{0})
  \label{eq:uniform-fit-budget}
\end{equation}
to measure the model's sensitivity to this error magnitude, and use
$\Delta\mathrm{PPL}_{\mathrm{inj}}(\tau)\leq\delta_{\mathrm{PPL}}$ as the search criterion. Previous work uses
similar plaintext error simulation to show that, when $\tau_0=2^{-12}$, the perplexity of Llama does not degrade
relative to FP16 inference~\cite{park2026llama}. We use $\tau_0$ as a conservative reference and comparison
baseline, rather than as a mandatory starting point, and test the error magnitude that the model can tolerate under
a given $\delta_{\mathrm{PPL}}$.

Directly generating 32 layers of polynomials for every candidate tolerance requires repeated Remez approximation
and full-model evaluation and is computationally expensive. We therefore replace polynomial generation in each
round with bounded perturbations that follow the candidate maximum-error range, simplifying the search from
``generate polynomials and evaluate'' to ``inject errors and evaluate.'' Error injection compares the relative
sensitivity of the model to different operators and error magnitudes. It does not reproduce the deterministic
residuals of a particular Remez polynomial, nor does it guarantee that the two produce the same PPL. The search
results are used only to select candidate budgets; model quality must still be evaluated separately after the final
polynomials are generated.

Specifically, we inject errors at the RMSNorm $\mathrm{rsqrt}$ outputs, Softmax $e^x$ outputs, and SiLU outputs
in all 32 layers. Because the Softmax auxiliary track necessarily undergoes BS during computation, the
inverse-square-root approximations for $j_1/j_2$ retain 12-bit precision and are excluded from tolerance
relaxation.

We first let the four nonlinear classes share a common tolerance $\tau_c$ and search, under different
$\delta_{\mathrm{PPL}}$ values, for the maximum value satisfying the above injection-test criterion.
Table~\ref{tab:common-fit-tolerance} reports the search results for the current 32-layer model. Even when the
perplexity increase in the injection test is limited to $0.001$, the candidate common tolerance reaches
$2.258\times10^{-3}$, which is $9.25$ times $2^{-12}$. Under the current experimental protocol, this result
indicates that the 12-bit reference configuration is conservative for bounded perturbations and motivates further
testing of operator-level error budgets.

\begin{table}[H]
  \centering
  \caption{Common approximation tolerances under different injection-test thresholds. All 32 layers share
  $\tau_c$; the injection positions for RMSNorm, Softmax, and SiLU are the $\mathrm{rsqrt}$ output, $e^x$
  output, and SiLU output, respectively.}
  \label{tab:common-fit-tolerance}
  \scriptsize
  \setlength{\tabcolsep}{3pt}
  \begin{tabular}{cccc}
    \toprule
    $\delta_{\mathrm{PPL}}$ & $\tau_c$ & vs. $2^{-12}$ & Measured $\Delta\mathrm{PPL}_{\mathrm{inj}}$ \\
    \midrule
    $0.001$ & $2.258\times10^{-3}$ & $9.25\times$ & $-0.00161$ \\
    $0.005$ & $2.354\times10^{-3}$ & $9.64\times$ & $0.00455$ \\
    $0.010$ & $2.403\times10^{-3}$ & $9.84\times$ & $0.00798$ \\
    $0.050$ & $2.838\times10^{-3}$ & $11.62\times$ & $0.04732$ \\
    \bottomrule
  \end{tabular}
\end{table}

\subsubsection{Cost-Aware Joint Error Allocation}
The common tolerance confirms that the 12-bit configuration can be relaxed, but does not exploit differences in
error sensitivity among operators. Let $\boldsymbol{\tau}$ denote the tolerance vector of the four nonlinear
classes. The $\tau_c\boldsymbol{1}$ obtained in the previous section is a common starting point that satisfies the
injection-test criterion. To refine it for different operators, we fix the current budgets of all other operators
and relax only operator $i$. We define the maximum tested value that satisfies the joint-injection criterion as the
conditional upper bound
\begin{equation}
  \widehat{\tau}_i(\boldsymbol{\tau}_{-i})=
  \max\left\{t\in[\tau_i,c_i]\mid
  \Delta\mathrm{PPL}_{\mathrm{inj}}(\boldsymbol{\tau}_{-i},t)
  \leq\delta_{\mathrm{PPL}}\right\},
  \label{eq:conditional-fit-budget}
\end{equation}
where $\boldsymbol{\tau}_{-i}$ denotes the current budgets excluding operator $i$, and $c_i$ is a predetermined
upper bound for the search. This conditional upper bound characterizes the error that can still be allocated to
operator $i$ under the current joint configuration, rather than an independent upper bound measured while all
other operators remain exact.

Algorithm~\ref{alg:joint-error-allocation} directly takes the $\tau_c$ obtained in the previous section as input.
It first initializes all four operator classes to $\tau_c$, then evaluates
Equation~\eqref{eq:conditional-fit-budget} coordinate by coordinate according to cost priority. Each candidate
value is evaluated over the complete model while perturbations in the other three classes are also present. Each
update therefore becomes the joint starting point for the next search. The entire procedure uses only the bounded
uniform-perturbation proxy; the polynomials corresponding to the final budgets are generated in
Section~\ref{sec:poly-generation}.

\begin{algorithm}[H]
  \caption{Cost-aware joint allocation of approximation errors}
  \label{alg:joint-error-allocation}
  \begin{algorithmic}[1]
    \Require Injection-test threshold $\delta_{\mathrm{PPL}}$, common tolerance $\tau_c$, operator search
    upper bounds $c_i$, priority order
    $\pi=[\mathrm{SiLU},\mathrm{RMSNorm2},\mathrm{RMSNorm1},\mathrm{Softmax}]$, and search precision $\epsilon$
    \Ensure Candidate joint perturbation budget $\boldsymbol{\tau}$
    \State $\tau_i\gets\tau_c$ for all operators $i$
    \For{$i$ \textbf{in} $\pi$}
      \State $l\gets\tau_i$, $r\gets c_i$
      \While{$r-l>\epsilon$}
        \State $m\gets(l+r)/2$; $\boldsymbol{\tau}'\gets\boldsymbol{\tau}$; $\tau_i'\gets m$
        \If{$\Delta\mathrm{PPL}_{\mathrm{inj}}(\boldsymbol{\tau}')\leq\delta_{\mathrm{PPL}}$}
          \State $l\gets m$
        \Else
          \State $r\gets m$
        \EndIf
      \EndWhile
      \State $\tau_i\gets l$
    \EndFor
    \State \Return $\boldsymbol{\tau}$
  \end{algorithmic}
\end{algorithm}

Because errors from the four operator classes propagate through residual connections, normalization, and
subsequent layers, the value of Equation~\eqref{eq:conditional-fit-budget} depends on the current
$\boldsymbol{\tau}_{-i}$. The conditional upper bounds of the operators therefore cannot be measured independently
and then directly combined. The error-allocation result of the operator-wise search also depends on the priority
order $\pi$.

The current strategy first relaxes SiLU. SiLU operates on the FFN intermediate state of width $14336$, and its
polynomial depth also affects the number of limbs that must be carried by the subsequent wide linear layer. When
tolerance relaxation causes the polynomial to cross a multiplicative-depth boundary, it can reduce the costs of
both nonlinear evaluation and adjacent linear computation. In contrast, a reduction in the degree of an RMSNorm
polynomial does not necessarily change the BS positions in the current execution graph. We therefore search SiLU,
RMSNorm2, RMSNorm1, and Softmax in this order.

Table~\ref{tab:joint-fit-budget} reports the joint-search result for $\delta_{\mathrm{PPL}}=0.001$. The common
starting point is $\tau_c=2.258\times10^{-3}$. The coordinate-wise search successively relaxes SiLU, RMSNorm2,
RMSNorm1, and Softmax to the conditional upper bounds shown in the table.
\begin{table}[H]
  \centering
  \caption{Candidate joint approximation-error budget under an injection-test threshold of
  $\delta_{\mathrm{PPL}}=0.001$.}
  \label{tab:joint-fit-budget}
  \scriptsize
  \setlength{\tabcolsep}{3pt}
  \begin{tabular}{lccc}
    \toprule
    Operator & Injection site & Tolerance $\tau_i$ & vs. $2^{-12}$ \\
    \midrule
    RMSNorm1 & $\mathrm{rsqrt}$ output & $9.881\times10^{-3}$ & $40.47\times$ \\
    Softmax & $e^x$ output & $3.093\times10^{-3}$ & $12.67\times$ \\
    RMSNorm2 & $\mathrm{rsqrt}$ output & $9.881\times10^{-3}$ & $40.47\times$ \\
    SiLU & SiLU output & $1.993\times10^{-2}$ & $81.65\times$ \\
    \bottomrule
  \end{tabular}
\end{table}

\subsubsection{Polynomial-Coefficient Generation}
\label{sec:poly-generation}
The operator-level budgets obtained from the joint search can be used directly as the maximum-error targets for
the corresponding approximation positions. For layer $l$ of operator $i$, the coefficient generator reads the
input interval $I_{i,l}$ determined in Section~\ref{sec:model-preprocess} and solves
\begin{equation}
  d_{i,l}^{*}=\min\left\{d\mid
  \max_{x\in I_{i,l}}\left|f_i(x)-p_{i,l,d}(x)\right|\leq\tau_i\right\},
  \label{eq:operator-approximation-target}
\end{equation}
Here, $\tau_i$ is the operator budget given in Table~\ref{tab:joint-fit-budget}, and $p_{i,l,d}$ is a degree-$d$
polynomial generated over the interval $I_{i,l}$. Because this budget is derived from an error-perturbation proxy,
the generated polynomials must still be verified using complete operator outputs and model-perplexity evaluation.

Given an interval and degree, the Remez algorithm generates an approximate minimax polynomial, but the required
degree cannot be directly derived from a target error. We therefore search for $d$ outside the Remez algorithm:
we progressively increase the candidate degree and check the maximum absolute error at the extrema. After
Equation~\eqref{eq:operator-approximation-target} is satisfied for the first time, we additionally check adjacent
lower degrees and finally determine $d_{i,l}^{*}$. For each operator, layer, and Softmax substep, the generated
coefficient package records the target function, approximation interval, error threshold, actual degree, and
coefficients for use during FHE execution.

Over the current approximation intervals for all 32 layers, the joint budget reduces the sum of SiLU polynomial
degrees from $2048$ to $736$ relative to a uniform 12-bit error target. It also reduces the sum of multiplicative
depths under the current evaluation strategy from $199$ to $155$. This result shows that relaxing model-level
error budgets can translate into reductions in polynomial degree and evaluation depth. Section~\ref{sec:evaluation}
evaluates its effects on operator latency and end-to-end performance.

\subsubsection{Treatment of CKKS Noise}

The preceding budget search considers only polynomial approximation errors and does not treat CKKS numerical
error as an independent search variable. Pilot tests show that, under the current parameters and ordinary signal
ranges, the absolute error of a single operation at a regular bootstrapping position is approximately $10^{-4}$
(and can be as low as $10^{-6}$ under a higher-precision configuration). This is lower than the
$10^{-3}$--$2\times10^{-2}$ approximation budgets obtained by the current joint search. This single-point result
indicates only the magnitude of CKKS error in the tested configuration and cannot replace an evaluation of
cumulative error over the complete model. We therefore do not include it in the model-level budget search at
present. Instead, Section~\ref{sec:eval-correctness} separately measures its incremental effect through the
difference between Poly-only and FHE execution. The scheduler in Section~\ref{sec:fhe-graph} determines the
specific bootstrapping positions and levels.

\subsection{Themis Packing Runtime}
\label{sec:linear-layout}

\subsection{Assembly and Scheduling of an FHE Transformer Layer}
\label{sec:fhe-graph}


\subsubsection{From Operator Contracts to a Single-Layer Computation Graph}

\subsubsection{Limb-Budget-Based Bootstrapping Schedule}
\section{Evaluation}
\label{sec:evaluation}

\subsection{Experimental Setup}
\label{sec:eval-setup}

All experiments are conducted on a server equipped with an AMD Ryzen 9 9950X CPU, 128~GB of host memory,
and one NVIDIA H100 80~GB GPU. The open-source EasyFHE library provides the underlying CKKS primitives and
GPU execution kernels. The model-conversion and experiment drivers are written in Python; PyTorch is used for
tensor preparation and reference evaluation, while the Themis packer invokes dedicated GPU kernels. We use
Meta-Llama-3-8B with a default input length of $S=128$. The CKKS ring dimension is $N=2^{16}$, the modulus
chain has a depth of 37, the first and remaining moduli are 60 bits and 59 bits, respectively, and $dnum=5$. THOR
and Themis use the same model weights, CKKS parameters, and bootstrapping configuration; each scheme uses a linear
data path and computational kernels matched to its packing scheme.

\subsection{Linear-Computation Performance}
\label{sec:eval-linear}
This section compares the performance of THOR and Themis on the main linear operators in Llama-3.
Table~\ref{tab:linear-performance} reports plaintext--ciphertext projections, including the QKV, $W_O$, gate/up,
and down projections. Table~\ref{tab:ccmm-performance} reports ciphertext--ciphertext computations, including
$QK^\top$ and $PV$ in attention and the SwiGLU gating product.

We further divide the online latency of plaintext--ciphertext projections into packing/encoding, homomorphic
evaluation, and rescaling. Both sets of experiments measure Layer~0 with the same model, input, and CKKS
parameters. THOR and Themis use computational kernels matched to their respective packing schemes. The results
therefore reflect the overall performance of each packing scheme and its corresponding linear implementation,
rather than only the speedup due to encoding reuse.

\begin{table}[H]
  \centering
  \caption{Online latency of THOR and Themis for plaintext--ciphertext projections in Layer~0}
  \label{tab:linear-performance}
  \scriptsize
  \setlength{\tabcolsep}{2.3pt}
  \resizebox{\columnwidth}{!}{%
  \begin{tabular}{lrrrrrrrrr}
    \toprule
    & \multicolumn{2}{c}{Pack/Encode (s)} & \multicolumn{2}{c}{Evaluation (s)} &
      \multicolumn{2}{c}{Rescale (s)} & \multicolumn{2}{c}{Total (s)} & Speedup \\
    \cmidrule(lr){2-3} \cmidrule(lr){4-5} \cmidrule(lr){6-7} \cmidrule(lr){8-9}
    Projection & THOR & Themis & THOR & Themis & THOR & Themis & THOR & Themis & \\
    \midrule
    QKV     & 6.72 & 0.18 & 3.62 & 0.91 & 0.16 & 0.00 & 10.50 & 1.10 & $9.5\times$ \\
    $W_O$   & 2.45 & 0.06 & 1.86 & 0.17 & 0.19 & 0.00 & 4.50 & 0.24 & $18.8\times$ \\
    gate/up & 14.83 & 0.55 & 3.90 & 2.47 & 0.52 & 0.01 & 19.24 & 3.02 & $6.4\times$ \\
    down    & 6.33 & 0.28 & 0.66 & 0.69 & 0.14 & 0.00 & 7.13 & 0.97 & $7.4\times$ \\
    \midrule
    Total   & 30.33 & 1.07 & 10.03 & 4.25 & 1.00 & 0.01 & 41.37 & 5.32 & $7.8\times$ \\
    \bottomrule
  \end{tabular}
  }%
\end{table}

In Table~\ref{tab:linear-performance}, the total time for plaintext--ciphertext computation decreases from
$41.37$~s with THOR to $5.32$~s with Themis, corresponding to a $7.8\times$ speedup. The original Pack/Encode
bottleneck decreases from $30.33$~s to $1.07$~s ($28.3\times$), and its proportion of the total decreases from
$73\%$ to $20\%$. Homomorphic-evaluation time decreases from $10.03$~s to $4.25$~s ($2.4\times$). In terms
of static operation counts, Themis reduces the number of weight encodings for QKV and gate/up to $1/16$ of that
required by diagonal packing without reuse, and reduces that for $W_O$ and down to $1/8$. These ratios compare
Themis with the no-reuse diagonal-packing formula as a static reference, rather than with THOR's measured Encode
invocations; they should not be interpreted as an isolated end-to-end THOR--Themis speedup.

In the ciphertext--ciphertext path, Themis uses a $C$--$\Delta$ layout designed for attention access patterns. This
layout pre-aligns $Q$ and $K/V$, allowing $QK^\top$ to directly produce $S_\Delta$ and allowing $PV$ to directly
consume the Softmax output $P_\Delta$. It thereby reduces the rotations required for ciphertext rearrangement and
eliminates intermediate repacking between the two kernels. Because the layout and the corresponding computational
kernels jointly determine data access and homomorphic operations, we report the complete computation path formed
by both and do not attribute the observed speedup to either factor alone.

\begin{table}[H]
  \centering
  \caption{Latency of THOR and Themis for ciphertext--ciphertext operations.
  $\operatorname{SiLU}(g)\odot u$ includes only element-wise multiplication and excludes SiLU polynomial evaluation.}
  \label{tab:ccmm-performance}
  \scriptsize
  \setlength{\tabcolsep}{4pt}
  \begin{tabular}{lrrr}
    \toprule
    Operator & THOR (s) & Themis (s) & Speedup \\
    \midrule
    $QK^\top$ & 9.88 & 0.34 & $29.1\times$ \\
    $PV$       & 2.38 & 0.46 & $5.2\times$ \\
    $\operatorname{SiLU}(g)\odot u$ & 0.04 & 0.02 & $2.1\times$ \\
    \midrule
    Total      & 12.30 & 0.82 & $15.0\times$ \\
    \bottomrule
  \end{tabular}
\end{table}

In addition to the linear-computation time reported in the table, THOR uses $0.54$~s to adjust the attention input
layout and $0.85$~s to repack the diagonal Softmax output into the probability copies required by $PV$. Themis's
$S_\Delta\rightarrow P_\Delta\rightarrow PV$ path eliminates the latter intermediate repacking while retaining the
QKV input adaptation and the PairFM conversion at the attention output. This layout difference and the associated
kernels jointly constitute the execution paths compared in Table~\ref{tab:ccmm-performance}.

\subsection{Ablation of Nonlinear Precision Allocation}
\label{sec:eval-nonlinear}

We use two progressive ablations to evaluate the computational costs of the candidate error budgets obtained in
Section~\ref{sec:nonlinear-calibration}. The first tests whether replacing uniform 12-bit precision with a common
tolerance $\tau_c$ reduces polynomial degree and multiplicative depth. Under the same injection-test threshold,
the second tests whether joint operator-wise allocation of $\boldsymbol{\tau}$ further reduces computation over
the common-tolerance configuration. We compare three configurations: uniform 12-bit precision, a common
tolerance, and joint allocation. All three use the same QuaRot weights, layer-wise approximation intervals, and
Remez coefficient-generation procedure, changing only the approximation-error budgets. Softmax-rsqrt is fixed at
12-bit precision in all configurations.

\begin{table}[H]
  \centering
  \caption{Nonlinear-polynomial costs under three error budgets}
  \label{tab:nonlinear-degrees}
  \scriptsize
  \setlength{\tabcolsep}{1.8pt}
  \resizebox{\columnwidth}{!}{%
  \begin{tabular}{lccccccccc}
    \toprule
    & \multicolumn{3}{c}{Uniform 12-bit} & \multicolumn{3}{c}{Common tolerance} &
      \multicolumn{3}{c}{Joint allocation} \\
    \cmidrule(lr){2-4}\cmidrule(lr){5-7}\cmidrule(lr){8-10}
    Operator & $\tau_i$ & $\sum d$ & $\sum L$ & $\tau_i$ & $\sum d$ & $\sum L$ &
      $\tau_i$ & $\sum d$ & $\sum L$ \\
    \midrule
    RMSNorm$_1$ & $2^{-12}$ & 8,032 & 269 & $\tau_c$ & 8,011 & 267 & $9.881\times10^{-3}$ & 7,962 & 265 \\
    RMSNorm$_2$ & $2^{-12}$ & 6,848 & 269 & $\tau_c$ & 6,806 & 267 & $9.881\times10^{-3}$ & 6,741 & 264 \\
    SiLU & $2^{-12}$ & \textbf{2,048} & \textbf{199} & $\tau_c$ & \textbf{1,076} & \textbf{174} & $\mathbf{1.993\times10^{-2}}$ & \textbf{736} & \textbf{155} \\
    Softmax-$e^x$ & $2^{-12}$ & \textbf{544} & \textbf{160} & $\tau_c$ & \textbf{288} & \textbf{127} & $\mathbf{3.093\times10^{-3}}$ & \textbf{279} & \textbf{127} \\
    Softmax-rsqrt & $2^{-12}$ & 6,976 & 479 & $2^{-12}$ & 6,976 & 479 & $2^{-12}$ & 6,976 & 479 \\
    \midrule
    Total & -- & 24,448 & 1,376 & -- & 23,157 & 1,314 & -- & 22,694 & 1,290 \\
    \bottomrule
  \end{tabular}
  }%
\end{table}

Table~\ref{tab:nonlinear-degrees} reports the actual polynomial costs generated by the three configurations.
$\sum d$ and $\sum L$ denote the sums of polynomial degrees and multiplicative depths, respectively, over 32
layers. After uniform 12-bit precision is relaxed to $\tau_c=2.258\times10^{-3}$, the total degree of all nonlinear
polynomials decreases from 24,448 to 23,157, and the total multiplicative depth decreases from 1,376 to 1,314.
With joint allocation, the total degree further decreases from 23,157 to 22,694, and the total depth decreases
from 1,314 to 1,290. For the highest-priority operator, SiLU, the total degree over 32 layers decreases from 2,048
to 736, and the total depth decreases from 199 to 155.

To quantify the observed runtime trend associated with these reductions in degree and depth, we use the polynomials
generated under the three error budgets and measure the online computation times of each nonlinear stage and the
complete FFN with the same input and CKKS parameters.
\begin{table}[H]
  \centering
  \caption{Online time of each computation stage under three error budgets with Themis packing}
  \label{tab:nonlinear-ablation}
  \scriptsize
  \setlength{\tabcolsep}{3pt}
  \begin{tabular}{lrrr}
    \toprule
    Stage & Uniform 12-bit (s) & Common (s) & Joint (s) \\
    \midrule
    RMSNorm$_1$ & 0.01 & 0.01 & 0.00 \\
    Softmax-$e^x$ & \textbf{0.11} & \textbf{0.07} & \textbf{0.07} \\
    Softmax-$j_1$ & 0.01 & 0.01 & 0.01 \\
    Softmax-$j_2$ & 0.04 & 0.05 & 0.05 \\
    RMSNorm$_2$ & 0.01 & 0.01 & 0.01 \\
    SiLU & \textbf{0.36} & \textbf{0.22} & \textbf{0.15} \\
    \midrule
    Tunable stages & 0.49 & 0.30 & 0.23 \\
    \bottomrule
  \end{tabular}
\end{table}

Table~\ref{tab:nonlinear-ablation} presents the two progressive comparisons in terms of runtime. The total time
of the four nonlinear stages participating in error allocation decreases from 0.49~s under uniform 12-bit
precision to 0.30~s under the common tolerance, and further to 0.23~s under joint allocation. The runtimes of
Softmax-$j_1$ and Softmax-$j_2$, which retain 12-bit precision, remain nearly unchanged across the three
configurations. The reduction in SiLU degree also reduces the FFN execution cost: the complete FFN time decreases
from 4.30~s to 4.05~s and then to 3.98~s. These measurements show the corresponding runtime reductions under
the tested configuration; they are not intended to characterize run-to-run variance.

\subsection{Single-Layer and Full-Model Performance}
\label{sec:eval-performance}

We measure steady-state performance for a representative intermediate layer. Table~\ref{tab:layer-performance}
reports the runtime of each main stage. The same nonlinear-polynomial configuration and bootstrapping policy are
used when profiling both schemes. The breakdown therefore includes the linear projections, layout conversions,
and implementation-dependent auxiliary work; it is a full-path comparison rather than an attribution of the total
difference to one stage. Themis reduces the single-layer linear stage from 38.38~s to 5.65~s and the complete-layer
time from 60.87~s to 11.12~s.

\begin{table}[H]
  \centering
  \caption{Runtime breakdown of Layer~16}
  \label{tab:layer-performance}
  \scriptsize
  \setlength{\tabcolsep}{3pt}
  \begin{tabular}{lrrr}
    \toprule
    Stage & THOR (s) & Themis (s) & Speedup \\
    \midrule
    RMSNorm                  & 0.22 & 0.49 & $0.4\times$ \\
    QKV projection           & 5.71 & 1.09 & $5.2\times$ \\
    QK--Softmax--PV          & 11.70 & 1.40 & $8.4\times$ \\
    $W_O$ projection         & 5.42 & 0.24 & $22.6\times$ \\
    gate/up--SiLU--down      & 27.24 & 4.32 & $6.3\times$ \\
    Layout conversion        & 0.92 & 1.27 & $0.7\times$ \\
    Bootstrap                & 5.10 & 2.30 & $2.2\times$ \\
    Other                    & 4.56 & 0.01 & -- \\
    \midrule
    Total                    & 60.87 & 11.12 & $5.5\times$ \\
    \bottomrule
  \end{tabular}
\end{table}

To extend from one layer to the complete model, the system connects 32 Transformer layers in sequence. Each layer
receives and outputs FM ciphertexts, so no additional layout conversion is required between layers. After the MLP
residual of a layer is completed, the system performs inter-layer bootstrapping to restore the limb budget required
by the RMSNorm and subsequent computation in the next layer.

Each layer uses the number of Softmax iterations $k$ determined during calibration, so the computational depths
are not identical across layers. Most intermediate layers use $k=2$; Layers~0--1 use $k=4$, Layers~2 and 14 use
$k=3$, and Layer~31 uses $k=5$. A larger $k$ requires more Softmax iterations and consumes additional depth, so
the system adds bootstrapping operations within these layers.

We refer to the consecutive server-side FHE evaluation of all 32 Transformer layers as end-to-end online
computation. With the same model, 128-token input, CKKS parameters, and hardware configuration, the end-to-end
online time is $1651.9$~s for THOR and $366.4$~s for Themis, corresponding to a $4.51\times$ speedup.

\subsection{Correctness and Model Quality}
\label{sec:eval-correctness}
This section evaluates model quality after Llama-3 is converted into an FHE computation graph. We compare Exact,
Poly-only, and FHE execution on the same model and inputs. Exact directly executes the original model and serves
as the quality baseline. Poly-only replaces nonlinear functions with fixed Remez polynomials and executes them in
plaintext to measure the quality loss caused by polynomial approximation; we report coefficients for both uniform
12-bit precision and joint allocation. FHE performs complete CKKS inference with the same polynomial
configuration to further measure the errors introduced by homomorphic operations and bootstrapping.

\begin{table}[H]
  \centering
  \caption{Model quality under different polynomial configurations and execution paths.}
  \label{tab:model-quality}
  \scriptsize
  \setlength{\tabcolsep}{5pt}
  \begin{tabular}{lrrr}
    \toprule
    Execution path & NLL & PPL & $\Delta$PPL (vs. Exact) \\
    \midrule
    Exact & 2.539 & 12.67 & -- \\
    Poly-only 12-bit & 2.549 & 12.79 & 0.122 \\
    Poly-only joint allocation & 2.549 & 12.79 & 0.119 \\
    FHE       & XXX & XXX & XXX \\
    \bottomrule
  \end{tabular}
\end{table}

We compute perplexity over eight WikiText-2 samples of length $128$, comprising $1016$ evaluation tokens. Exact
has a PPL of $12.67$. The uniform 12-bit and joint-allocation Poly-only paths both have a PPL of $12.79$, with
$\Delta$PPL values of $0.122$ and $0.119$ relative to Exact, respectively. This result shows that after relaxing
approximation tolerances by operator, the plaintext-polynomial quality is comparable to that of the 12-bit
baseline. This result directly evaluates the fixed Remez polynomials. Error injection is used only to compare
sensitivity and select candidate budgets, so its $\Delta$PPL need not numerically match the $\Delta$PPL of the
final polynomials. With the same polynomial configuration, the final FHE PPL is XXX, a further change of XXX
relative to Poly-only; this difference measures the incremental error introduced by CKKS operations and
bootstrapping. In a separate 128-token prefill regression case, all three FHE repetitions decode the expected next
token.

\subsection{Comparison with Existing FHE LLM Systems}
\label{sec:eval-related}

Table~\ref{tab:system-comparison} compares Themis with representative non-interactive FHE Transformer/LLM
systems. THOR and MOAI evaluate BERT-base, whereas Scaling up Privacy-Preserving ML evaluates Llama-2-7B and
encrypts the final 128 tokens of a 4096-token input. Because the models, encrypted-input ranges, hardware counts,
and timing boundaries differ, the values in the table illustrate implementation scale and execution protocols and
are not used to compute cross-paper speedups. In comparison, Themis evaluates a 128-token prefill protocol for
Llama-3-8B on a single H100 with an online latency of $366.4$~s.

\begin{table}[H]
  \centering
  \caption{Comparison with existing FHE Transformer/LLM systems. Latencies are not directly comparable because
  the models, encrypted ranges, and hardware differ.}
  \label{tab:system-comparison}
  \scriptsize
  \setlength{\tabcolsep}{2.5pt}
  \begin{tabular}{lp{0.17\columnwidth}p{0.27\columnwidth}p{0.18\columnwidth}r}
    \toprule
    System & Model & Input protocol & Hardware & Latency \\
    \midrule
    THOR~\cite{thor} & BERT-base & 128 encrypted & 1$\times$A100 & 602.26~s \\
    MOAI~\cite{moai2025} & BERT-base & 128 encrypted & 1$\times$H200 & 141.3~s \\
    Scaling~\cite{park2026llama} & Llama-2-7B & 4096 total, 128 encrypted & 8$\times$RTX~4090 & 85~s \\
    \sysname{} & Llama-3-8B & 128-token Prefill & 1$\times$H100 & $366.4$~s \\
    \bottomrule
  \end{tabular}
\end{table}

\section{Conclusion}
\label{sec:conclusion}

We present \sysname{}, a non-interactive FHE inference system for Llama-3. To address the large amount of
redundant weight encoding produced by THOR-style packing in wide linear projections, Themis uses a feature-major
layout to unify residual and inter-layer interfaces and constructs short-lived specialized layouts for linear
projections and attention. This design reuses encoded weight diagonals. Within attention, $QK^\top$ produces
scores that Softmax can consume directly, and $PV$ directly consumes the resulting probabilities. The system
further combines input-interval calibration, minimax polynomial approximation, and model-level joint error
allocation to implement RMSNorm,
Softmax, and SiLU, and assembles these operators into a complete CKKS computation graph for Llama-3.

For Llama-3-8B with a 128-token input on a single NVIDIA H100 80~GB GPU, Themis reduces the single-layer
plaintext--ciphertext projection time from $41.37$~s to $5.32$~s and the execution time of an intermediate layer from
$60.87$~s to $11.12$~s. The online time for the complete 32-layer ciphertext computation decreases from
$1651.9$~s with THOR to $366.4$~s with Themis, corresponding to a $4.51\times$ speedup, with peak device memory
usage of $58.9$~GiB. These results show that the packing scheme for an FHE LLM determines not only the operation
counts of an individual matrix multiplication, but also whether weight encoding, inter-operator representations,
and layer-level scheduling can form an efficient complete execution path. Future work can extend this cross-layer
interface to more models and input scales and further co-optimize packing, nonlinear precision, and bootstrapping
schedules.

\bibliographystyle{plain}
\bibliography{refs}

@inproceedings{cheon2017ckks,
  title     = {Homomorphic Encryption for Arithmetic of Approximate Numbers},
  author    = {Cheon, Jung Hee and Kim, Andrey and Kim, Miran and Song, Yongsoo},
  booktitle = {Advances in Cryptology -- ASIACRYPT 2017},
  year      = {2017}
}

@book{devore1993constructive,
  title     = {Constructive Approximation},
  author    = {DeVore, Ronald A. and Lorentz, George G.},
  series    = {Grundlehren der mathematischen Wissenschaften},
  volume    = {303},
  publisher = {Springer-Verlag},
  year      = {1993}
}

@inproceedings{vaswani2017attention,
  title     = {Attention Is All You Need},
  author    = {Vaswani, Ashish and Shazeer, Noam and Parmar, Niki and Uszkoreit, Jakob and Jones, Llion and Gomez, Aidan N. and Kaiser, Lukasz and Polosukhin, Illia},
  booktitle = {Advances in Neural Information Processing Systems},
  year      = {2017}
}

@misc{dubey2024llama3,
  title         = {The {Llama} 3 Herd of Models},
  author        = {Dubey, Abhimanyu and Jauhri, Abhinav and Pandey, Abhinav and others},
  year          = {2024},
  eprint        = {2407.21783},
  archivePrefix = {arXiv},
  primaryClass  = {cs.AI}
}

@misc{park2026llama,
  title        = {Scaling up Privacy-Preserving {ML}: A {CKKS} Implementation of {Llama}-2-{7B}},
  author       = {Park, Jaiyoung and Park, Sejin and Park, Jai Hyun and Ahn, Jung Ho and Cheon, Jung Hee and Hanrot, Guillaume and Kim, Jung Woo and Park, Minje and Stehl{\'e}, Damien},
  year         = {2026},
  eprint       = {2601.18511},
  archivePrefix= {arXiv},
  primaryClass = {cs.CR},
  note         = {PDF: references/Park\_et\_al\_Llama2\_FHE\_arxiv2601.18511.pdf; complete author list}
}

@inproceedings{bae2024pcmm,
  title     = {Plaintext-Ciphertext Matrix Multiplication and {FHE} Bootstrapping: Fast and Fused},
  author    = {Bae, Youngjin and Cheon, Jung Hee and Hanrot, Guillaume and Park, Jai Hyun and Stehl{\'e}, Damien},
  booktitle = {Advances in Cryptology -- CRYPTO 2024},
  year      = {2024}
}

@misc{thor,
  title        = {{THOR}: Secure Transformer Inference with Homomorphic Encryption},
  author       = {Moon, Jungho and Yoo, Dongwoo and Jiang, Xiaoqian and Kim, Miran},
  year         = {2025},
  note         = {Local reference PDF; final publication metadata to verify before submission}
}

@misc{moai2025,
  title        = {{MOAI}: Module-Optimizing Architecture for Non-Interactive Secure Transformer Inference},
  author       = {Zhang, Linru and Wang, Xiangning and Sim, Jun Jie and Huang, Zhicong and Zhong, Jiahao and Wang, Huaxiong and Duan, Pu and Lam, Kwok-Yan},
  year         = {2025},
  note         = {Local reference PDF: references/MOAI.pdf; final publication metadata to verify before submission}
}

@inproceedings{ashkboos2024quarot,
  title     = {{QuaRot}: Outlier-Free {4}-Bit Inference with Rotation of {LLMs}},
  author    = {Ashkboos, Saleh and Mohtashami, Amirkeivan and Croci, Maximilian L. and others},
  booktitle = {NeurIPS},
  year      = {2024},
  note      = {PDF: references/QuaRot.pdf}
}

@inproceedings{xiao2024streamingllm,
  title     = {Efficient Streaming Language Models with Attention Sinks},
  author    = {Xiao, Guangxuan and Tian, Yuandong and Chen, Beidi and Han, Song and Lewis, Mike},
  booktitle = {International Conference on Learning Representations},
  year      = {2024}
}

@misc{hosomax,
  title        = {Fast and Accurate Homomorphic Softmax Evaluation},
  author       = {Cho, Wonhee and Hanrot, Guillaume and Kim, Taeseong and Park, Minje and Stehl{\'e}, Damien},
  year         = {2024},
  eprint       = {2410.11184},
  archivePrefix= {arXiv},
  primaryClass = {cs.CR}
}

@inproceedings{iron2022,
  title     = {{Iron}: Private Inference on Transformers},
  author    = {Hao, Meng and Li, Hongwei and Chen, Hanxiao and Xing, Pengzhi and Xu, Guowen and Zhang, Tianwei},
  booktitle = {Advances in Neural Information Processing Systems},
  volume    = {35},
  pages     = {15718--15731},
  year      = {2022}
}

@misc{puma2023,
  title         = {{PUMA}: Secure Inference of {LLaMA}-7B in Five Minutes},
  author        = {Dong, Ye and Lu, Wen-jie and Zheng, Yancheng and Wu, Haoqi and Zhao, Derun and Tan, Jin and Huang, Zhicong and Hong, Cheng and Wei, Tao and Chen, Wenguang},
  year          = {2023},
  eprint        = {2307.12533},
  archivePrefix = {arXiv},
  primaryClass  = {cs.CR}
}

@inproceedings{bolt2024,
  title     = {{BOLT}: Privacy-Preserving, Accurate and Efficient Inference for Transformers},
  author    = {Pang, Qi and Zhu, Jinhao and M{\"o}llering, Helen and Zheng, Wenting and Schneider, Thomas},
  booktitle = {2024 IEEE Symposium on Security and Privacy},
  pages     = {4753--4771},
  year      = {2024}
}

@inproceedings{bumblebee2025,
  title     = {{BumbleBee}: Secure Two-Party Inference Framework for Large Transformers},
  author    = {Lu, Wen-jie and Huang, Zhicong and Gu, Zhen and Li, Jingyu and Liu, Jian and Hong, Cheng and Ren, Kui and Wei, Tao and Chen, Wenguang},
  booktitle = {Network and Distributed System Security Symposium},
  year      = {2025}
}

@inproceedings{nexus2025,
  title     = {Secure Transformer Inference Made Non-Interactive},
  author    = {Zhang, Jiawen and Liu, Jian and Yang, Xinpeng and Wang, Yinghao and Chen, Kejia and Hou, Xiaoyang and Ren, Kui and Yang, Xiaohu},
  booktitle = {Network and Distributed System Security Symposium},
  year      = {2025}
}

\end{document}